\documentclass[aps,pra,twocolumn,showpacs]{revtex4}
\usepackage{graphicx}
\usepackage{amsmath,amsfonts}

\newcommand{\er}{\mathbf{r}}
\newcommand{\ee}{{\rm e}}

\begin{document}

\title{Ground states of trapped spin-1 condensates in magnetic field}

\author{Micha\l{} Matuszewski}
\affiliation{Instytut Fizyki PAN, Aleja Lotnik\'ow 32/46, 02-668 Warsaw, Poland}

\begin{abstract}
We consider a spin-1 Bose-Einstein condensate trapped in a harmonic potential under the influence of a homogeneous magnetic field.
We investigate spatial and spin structure of the mean-field ground states under 
constraints on the number of atoms and the total magnetization. 
We show that the trapping potential can
make the antiferromagnetic condensate separate into three, and ferromagnetic condensate into two distinct phases. 
In the ferromagnetic case, the magnetization is located in the center
of the harmonic trap, while in the antiferromagnetic case magnetized phases appear in the outer regions.
We describe how the transition from the Thomas-Fermi regime to the single-mode approximation regime with 
decreasing number of atoms results in the
disappearance of the domains. We suggest that the ground states can be created in experiment by adiabatically 
changing the magnetic field strength.
\end{abstract}

\maketitle

\section{Introduction}

Bose-Einstein condensates with spin degrees of freedom~\cite{Ho} attracted in recent years 
great interest due to the unique possibility of exploring fundamental concepts of quantum mechanics
in a remarkably controllable and tunable environment.
The ability to generate spin squeezing and entanglement~\cite{Entanglement}
makes spinor Bose gases promising candidates for 
applications as quantum simulators~\cite{QS}, in quantum information~\cite{QI}, and for precise measurements~\cite{Measurement}.
Moreover, spinor condensates were successfully used to recreate many of the phenomena of condensed matter physics
in experiments displaying an unprecedented level of control over the quantum system.
In particular, spin domains~\cite{Stenger_Nat_1998,Ketterle_Metastable,Sadler_Nat_2006}, 
spin mixing~\cite{Mixing,Chang_PRL_2004}, and spin vortices~\cite{Ketterle_Coreless}
were predicted and observed.

The ground states of spin-1 condensates in homogeneous magnetic field 
have been studied in a number of previous works \cite{Zhang_NJP_2003,Stenger_Nat_1998,Ueda_SBS,SMA,Zhou,Matuszewski_PS}.
The typical procedure consists of minimization of the total energy under constraints on the number of atoms and 
the longitudinal magnetization, which is well conserved in typical experimental conditions~\cite{Chang_PRL_2004}.
Most of the studies, however, considered the condensate in the single-mode approximation, which assumes that the 
spin components share the same spatial profile \cite{SMA,Zhang_PRA_2005}, ignoring the possibility of phase separation.

In Ref.~\cite{Zhang_NJP_2003}, the breakdown of the single-mode approximation was shown numerically for a condensate confined
in a harmonic potential. In a recent paper~\cite{Matuszewski_PS}, we studied the possibility of phase separation
in the case of an untrapped condensate, and demonstrated that this phenomenon can take place in an 
antiferromagnetic condensate, leading to the splitting into two distinct phases. On the other hand, ferromagnetic condensates
were found not to display phase separation for any values of parameters. However, the trapping potential,
present in any experimental realization, can strongly influence the structure of the ground states. Indeed, 
in the context of binary condensates, it was previously shown that spatial separation of components can be induced
by an external potential~\cite{Timmermans}. This phenomenon was called ``potential separation'' as opposed to 
``phase separation'' which occurs spontaneously also in the untrapped case. Moreover, the external potential
determines the spatial structure of phase separated states. In general, phases that are characterized by energetically
unfavorable interatomic interactions are moved towards the outer, low density regions, as demonstrated
in spin-imbalanced Fermi systems~\cite{Hulet_ImbalancedFermions} and spinful bosons~\cite{Cornell_BinarySeparation,Matuszewski_PS}.

In this paper, we investigate in detail the spatial separation of spin phases in the ground states of spin-1 condensates
trapped in harmonic potentials. 
We find that the phase diagram of a trapped condensate is substantially different from the one
of an untrapped condensate. We show that the external potential can
make the antiferromagnetic condensate separate into three, and ferromagnetic condensate into two distinct phases. 
Furthermore, we suggest an experimental method for the creation of these ground states 
by adiabatically changing the magnetic field strength. 

The paper is organized as follows. Section~\ref{Sec_model} reviews the mean-field model of a spin-1 condensate 
in a homogeneous magnetic field and the possible spin phases. 
Section~\ref{Sec_TF} presents analytical and qualitative results describing the structure of ground states in the Thomas-Fermi
approximation. In Section~\ref{Sec_Numerical} we treat the problem numerically in a systematic way
and demonstrate the crossover between the Thomas-Fermi and single-mode-approximation regimes.
Section~\ref{Sec_Generation} describes an experimental method of creation of the ground states.
Section~\ref{Sec_conclusions} concludes the paper.

\section{Model} \label{Sec_model}

We consider a dilute spin-1 BEC in a homogeneous magnetic field pointing along the $z$ axis.
We start with the mean-field Hamiltonian $H = H_{\rm S} + H_{\rm A}$,
\begin{equation} \label{En}
H = \sum_{j=-,0,+} \int d\er \, \psi_j^* \left(-\frac{\hbar^2}{2m}\nabla^{2} + \frac{c_0}{2} n 
+ V({\bf r})\right) \psi_j + H_A,
\end{equation}
where $\psi_-,\psi_0,\psi_+$ are the wavefunctions of atoms in magnetic sublevels $m_{\rm f}=-1,0,+1$, 
$M$ is the atomic mass, $V({\bf r})$ 
is an external potential and $n=\sum n_j = \sum |\psi_j|^2$ is the total atom density.  
The asymmetric (spin dependent) part of the Hamiltonian is given by
\begin{equation}
\label{EA}
H_{\rm A} = \int d\er \left(\sum_{j=-,0,+} E_jn_j + \frac{c_2}{2}|{\bf F}|^2\right)
\end{equation}
where $E_j$ is the Zeeman energy shift for state $\psi_j$ and the spin density is,
\begin{equation}
\label{spindensity}
{\bf F}=(F_x,F_y,F_z)=(\psi^{\dagger}\hat{F}_x\psi,\psi^{\dagger}\hat{F}_y\psi,\psi^{\dagger}\hat{F}_z\psi)
\end{equation}
where $\hat{F}_{x,y,z}$ are the spin-1 matrices~\cite{Isoshima_PRA_1999} and $\psi =(\psi_+,\psi_0,\psi_-)$.  
The spin-independent and spin-dependent interaction coefficients are given by $c_0=4
\pi \hbar^2(2 a_2 + a_0)/3m$ and $c_2=4 \pi \hbar^2(a_2 -
a_0)/3m$, where $a_S$ is the s-wave scattering length for colliding atoms
with total spin $S$.
The total number of atoms and the total magnetization in the direction of the magnetic field
\begin{align}
N&=\int n d \er\,, \\
\mathcal{M}&= \int F_z d \er = \int \left(n_+ -
n_-\right) d \er\,,
\end{align}
are conserved quantities.
The Zeeman energy shift for each of the components, $E_j$ can be calculated using the
Breit-Rabi formula~\cite{Wuster}
\begin{align}
E_{\pm}& = -\frac{1}{8}E_{\rm HFS}\left(1 + 4\sqrt{1\pm \alpha + \alpha^2} \right)  \mp g_I \mu_B B\,, \nonumber \\
E_{0} &= -\frac{1}{8}E_{\rm HFS}\left(1 + 4\sqrt{1 + \alpha^2} \right)\,,
\label{BR}
\end{align}
where $E_{\rm HFS}$ is the hyperfine energy splitting at zero
magnetic field, $\alpha = (g_I + g_J) \mu_B B/E_{\rm HFS}$, where
$\mu_B$ is the Bohr magneton, $g_I$ and $g_J$ are the gyromagnetic
ratios of nucleus and electron, and $B$ is the magnetic field strength.
The linear part of the Zeeman effect gives rise to an
overall shift of the energy, and so we can remove it with the
transformation
\begin{equation}
H \rightarrow H + (N + \mathcal{M}) E_+/2 + (N - \mathcal{M}) E_-/2\,.
\end{equation}
This transformation is equivalent to the removal of the Larmor precession
of the spin vector around the $z$ axis \cite{Matuszewski_PRA_2008,Ueda_SBS}.
We thus consider only the effects of the quadratic Zeeman shift.
For sufficiently weak magnetic field we can approximate it by $\delta
E=(E_+ + E_- - 2E_0)/2  \approx \alpha^2 E_{\rm HFS}/16$, which is always
positive.

The asymmetric part of the Hamiltonian (\ref{EA}) can now be rewritten as
\begin{equation} \label{EA2}
H_{\rm A} = \int d\er \, \left(-\delta E \,n_0 + \frac{c_2}{2} |{\bf F}|^2\right)= \int d\er \, n\, e(\er)\,,
\end{equation}
where the energy per atom $e(\er)$ is given by \cite{Zhang_PRA_2005}
\begin{align} \label{EA3}
e =& -\delta E \rho_0+\frac{c_2 n}{2} |\mathbf{f}|^2 = - \delta E \rho_0 + \frac{c_2 n}{2}\left( |{\bf f}_\perp|^2 +m^2\right)\,, \nonumber\\
|{\bf f}_\perp|^2 =& 2\rho_0 (1-\rho_0) + 2\rho_0 \sqrt{(1-\rho_0)^2-m^2} \cos(\theta)\,.
\end{align}
We express the wavefunctions as $\psi_j = \sqrt{n \rho_j} \exp(i \theta_j)$ where the relative densities are $\rho_j=n_j/n$.
We also introduced the relative phase $\theta = \theta_+ + \theta_- - 2\theta_0$,
spin per atom $\mathbf{f} = \mathbf{F}/n$, and magnetization per atom $m = f_z=\rho_+-\rho_-$.  The perpendicular spin component per atom is $|{\bf f}_\perp|^2 = f_x^2+f_y^2$.

The Hamiltonian~(\ref{En}) gives rise to the Gross-Pitaevskii equations describing the mean-field dynamics of the system
\begin{align}\label{GP}
i \hbar\frac{\partial \psi_{\pm}}{\partial t}&=\left[ \mathcal{L} +
c_2 (n_{\pm} + n_0 - n_{\mp})\right] \psi_{\pm} +
c_2 \psi_0^2 \psi_{\mp}^* \,, \\\nonumber
i \hbar\frac{\partial \psi_{0}}{\partial t}&=\left[ \mathcal{L} -
\delta E + c_2 (n_{+} + n_-)\right] \psi_{0} + 2 c_2
\psi_+ \psi_- \psi_{0}^* \,,
\end{align}
where $\mathcal{L}$ is given by $\mathcal{L}=-\hbar^2\nabla^2/2M+c_0n + V({\bf r})$.

By comparing the kinetic energy with the interaction energy, we can determine the
healing length $\xi=2\pi\hbar / \sqrt{2M c_0 n}$ and the spin healing length
$\xi_s=2\pi\hbar / \sqrt{2M c_2 n}$. These quantities give the length scales of
spatial variations in the condensate profile induced by the spin-independent or spin-dependent
interactions, respectively. Analogously, we define the magnetic healing length as
$\xi_B=2\pi\hbar / \sqrt{2M \delta E}$.

In spin-1 condensates created to date, the $a_0$ and $a_2$ scattering lengths have similar magnitudes.
The spin-dependent interaction coefficient $c_2$ is then much smaller than its spin-independent
counterpart $c_0$. For example, this ratio is about 1:30 in a $^{23}$Na condensate and 1:220 in a $^{87}$Rb condensate 
far from Feshbach resonances \cite{Beata}.
Consequently, changing the total density $n$ requires much more energy than changing the relative populations of spin states $n_j$.
In our considerations we will treat the total atom density profile $n(\er)$ as a constant, close to the Thomas-Fermi profile
for a given potential $V(\er)$.

\section{Homogeneous stationary states} \label{Sec_SS}

We recall the possible phases of spin-1 condensates in magnetic field~\cite{Matuszewski_PS}.
The stationary solutions in the case of a vanishing potential, $V({\bf r}) = 0$, have the form
\begin{equation}
\label{stat}
\psi_{j}(\er,t) = \sqrt{n_{j}} \ee^{-i(\mu_{j} + \mu_S)t + i \theta_j}\,,
\end{equation}
where $\mu_S=c_0 n / \hbar$ is a constant.
These solutions are stationary in the sense that the number of atoms in each magnetic sublevel $n_j$ is
constant in time, but the relative phases may change as a result of an additional spin precession around $z$,
as long as the phase matching condition
\begin{equation}
\mu_+ + \mu_- = 2\mu_0\,, \label{PM}
\end{equation}
is fulfilled \cite{Isoshima_PM,Matuszewski_PRA_2008}.
Because the symmetric part of the Hamiltonian in~(\ref{En}) is constant, the relevant part of the Hamiltonian is given by Eq.~(\ref{EA2}).

The Hamiltonian~(\ref{En}) and GP equations~(\ref{GP}) are invariant under the gauge transformation $\psi_j \rightarrow \psi_j \exp(-i \beta)$
and rotation around the $z$ axis $\psi_j \rightarrow \psi_j \exp(-i F_z \gamma)$, which transform the wavefunction components according to
\begin{equation}
\label{rotations}
\left( \begin{array}{c} \psi_+ \\ \psi_0 \\ \psi_- \end{array} \right) \rightarrow
\ee^{-i\beta}\left( \begin{array}{c} \ee^{-i\gamma} \psi_+ \\ \psi_0 \\ \ee^{i\gamma} \psi_- \end{array} \right) \,.
\end{equation}
Hence the solutions can be classified using the relative densities $\rho_j=n_j/n$ and 
a single relative phase $\theta=\theta_+ + \theta_- - 2\theta_0$,
with the chemical potentials $\mu_j$ given as solutions to Eqs.~(\ref{GP}).  We note that for stationary solutions 
with all three components populated the relative phase 
must take one of two values, $\theta = 0$ or $\theta = \pi$.  We call the former ``phase-matched" states (PM) and 
the latter ``anti-phase-matched" (APM) states.  The names derive from the fact that within the continuum of 
states satisfying the spin rotations (\ref{rotations}) there is a set $(\psi_+,\psi_0,\psi_-)$ with all 
components in phase for the PM states, and a set with $\psi_+$ and $\psi_0$ in phase but with $\psi_-$ $\pi$ out of phase 
for the APM states.

In general, the possible stationary states can be classified as follows
\begin{enumerate}
\item Nematic state ($\rho_0$), with all the atoms in the $m_f=0$ component, $\psi=(0,\psi_0,0)$. The chemical potential,
spin density and energy per atom are equal to
\begin{equation}
\hbar\mu_0 = -\delta E, \quad |{\bf f}|=0, \quad e = -\delta E\,.
\end{equation}
\item Magnetized states ($\rho_-$ and $\rho_+$), with all the atoms in the $m_f=-1$ or in the $m_f=+1$ component,
$\psi=(0,0,\psi_+)$ or $\psi=(\psi_-,0,0)$
\begin{equation}
\hbar \mu_\pm= c_2n, \quad |m|=1, \quad |{\bf f}_\perp|=0, \quad e = \frac{c}{2}\,.
\end{equation}
\item Two-component or stretched states (2C), with $\psi = (\psi_-, 0, \psi_+)$ and arbitrary magnetization
\begin{equation}
\hbar\mu_\pm = \pm c_2 n m, \quad |{\bf f}_\perp|=0, \quad e = \frac{c}{2} m^2\,.
\end{equation}
\item Phase-matched states (PM), where all three sublevels are populated, $\psi = (\psi_-, \psi_0, \psi_+)$, and
the relative phase is equal to $\theta=0$.
\item Anti-phase-matched states (APM), similar as the PM state but with $\theta=\pi$.
\end{enumerate}

The parameters of the PM and APM phases cannot be expressed by analytical formulas in the general case.
Since all three components are populated in these states, the magnetization perpendicular to the magnetic field
direction, $|{\bf F}_\perp|=\sqrt{2}|\psi_+\psi_0^*+\psi_0\psi_-^*|$, is nonzero. Hence the axial symmetry of
the system is broken~\cite{Ueda_SBS}.

These above classification is also applicable to inhomogeneous condensates
within the single-mode approximation (SMA), which assumes that the spin components share
the same spatial profile \cite{SMA,Zhang_PRA_2005}, after replacing $n$ with $\langle n \rangle$.
This assumption is true  eg.~when the condensate size is much smaller
than the spin healing length $\xi_s$ and the magnetic healing length $\xi_B$.

\section{Ground states in the Thomas-Fermi approximation}\label{Sec_TF}

In this section, we investigate the structure of the ground states in harmonic potential within the Thomas-Fermi (TF)
approximation (or local density approximation), neglecting the kinetic energy in the Hamiltonian~(\ref{En}).
This approximation is justified when the spatial variation of the condensate wavefunction gives a contribution 
that is relatively small, in particular when the size of the condensate is much larger than both $\xi_s$ and $\xi_B$. 

Within our assumptions, the asymmetric part of the energy~(\ref{EA}) is a small contribution to the total energy.
The profile of the total density is determined by the minimization of the symmetric part of the 
free energy with a fixed particle number
\begin{equation}
K_{\rm S}=\int \left(\frac{c_0}{2}n + V\right) n \,d\er - \mu N\,,
\end{equation}
where $\mu$ is the Lagrange multiplier corresponding to the chemical potential.
The minimization of this functional gives the well known Thomas-Fermi profile $c_0n(\er)+V(\er)=\mu={\rm const}$.

Analogously, to determine the spin structure, 
we minimize the asymmetric part of the energy $E_{\rm A}$ with a fixed total magnetization $M$
\begin{equation}
K_{\rm A}=\int e(n, m) n \,d\er - pM,\,\, M=\int m n \,d\er.
\end{equation}
The minimum of this functional requires that
\begin{equation}
\frac{\partial e(n,m)}{\partial m} = p=\mathrm{const},\label{dem}
\end{equation}
when moving from one point in space to another. This condition is not applicable to the completely magnetized $\rho_+$, $\rho_-$
and the unmagnetized $\rho_0$ phase, 
for which $\partial e/\partial m$ is undefined~\cite{Matuszewski_PS}.

\subsection{Antiferromagnetic case ($c_2 >0$)}

Since we work within the Thomas-Fermi approximation and neglect the spatial derivatives,
at each point in space the local phase corresponds to a 
ground state of a homogeneous condensate~\cite{Matuszewski_PS}. 
Nevertheless, the change of local parameters (density and magnetization)
in space may lead to the change of the local ground state and the appearance of spatial domains.

In the case of an antiferromagnetic condensate,
there are five possible local ground state phases: $\rho_+$,  $\rho_-$, $\rho_0$, 2C and APM~\cite{Matuszewski_PS}. 
The last two of them allow for a variation of the local magnetization. In the case of the
2C phase, the condition~(\ref{dem}) gives
\begin{equation} \label{2C_cond}
\frac{\partial e}{\partial m} = n m = n_+ - n_- = \mathrm{const},
\end{equation}
hence the difference in the density of atoms in the $m=+1$ and the $m=-1$ components is constant. For the APM state, 
the asymmetric energy (further on we call it simply the energy)
is not an analytical function of $n$ and $m$, and we approximate it with the formula 
$e_{\rm APM}\approx (|m|-1) \delta E + |m|c_2 n/2$, which gives less than $10\%$ error for any $n$ and $m$. 
The condition~(\ref{dem}) gives
\begin{equation} \label{APM_cond}
\frac{\partial e}{\partial m} = \delta E + \frac{c_2 n}{2} = \mathrm{const},
\end{equation}
which obviously cannot be fulfilled if the density profile is not homogeneous. Consequently, the APM phase is unstable, 
because the transfer of magnetization from the central, high density area to the outer regions is always energetically favorable.
Hence, the ground state can contain $\rho_+$,  $\rho_-$, $\rho_0$, and/or 2C domains, the latter fulfilling the 
condition~(\ref{2C_cond}). 

\begin{figure}
\includegraphics[width=6.5cm]{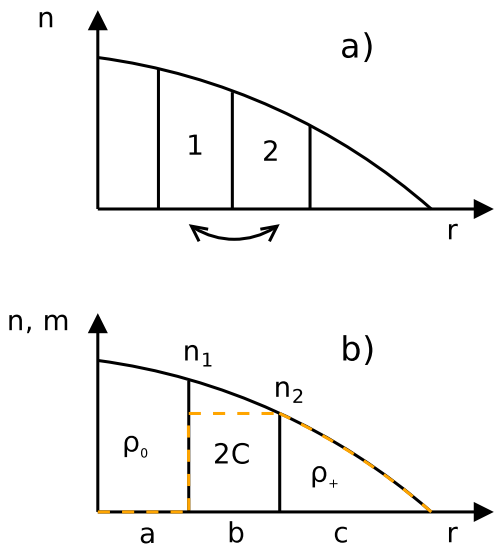}
\caption{Schematic profiles of phase separated states. (a) The swapping of two domains can change the total energy,
even if the number of atoms in each of them is kept constant. 
(b) The structure of the ground
state of an antiferromagnetic condensate composed of three phases. The dashed line shows the local magnetization.
}
\label{structure}
\end{figure}

The other question is the placement of the domains. Since the local ground state depends on the density,
the domains arrange according to the distance from the center of the trap, as in Fig.~\ref{structure}(a). Let us consider two domains,
one of them composed of phase X placed at the area 1 and the 
other composed of phase Y placed at the area 2, containing equal number of atoms. 
If the energy per atom of the phase X is constant, (which is the case for the
$\rho_0$ phase), then swapping of the domains as in Fig.~\ref{structure}(a) will not lead to the change of its energy 
\begin{equation}
\Delta E_X  = e_{\rm X} \left(\int_1 n \,d\er - \int_2 n \,d\er\right) = 0\,.
\end{equation}
On the other hand, if the energy per atom of the phase Y is proportional to the number of atoms, $e_{\rm Y} = a n$, 
as for the $\rho_\pm$ or 2C phases, the difference in its energy is 
\begin{equation}
\Delta E_Y  = a \left(\int_1 n^2 \,d\er - \int_2 n^2 \,d\er\right) = a B,\, \mathrm{where} \,B>0\,.
\end{equation}
In result, it is energetically favorable for the $\rho_0$ phase to stay in the high density area in the center of the trap.
Moreover, the $\rho_+$ or $\rho_-$ phase will be moved further than the 2C phase to the outer areas, 
since their coefficient $a$ is larger. 

We obtain the general domain structure depicted in Fig.~\ref{structure}(b), shown for positive $M$
(for negative $M$, the local magnetization is opposite and the $\rho_+$ state is replaced by  $\rho_-$).
The boundary between the phases $\rho_0$ and 2C corresponds to a first order phase transition, and the
boundary between the phases 2C and $\rho_+$ to a second order phase transition~\cite{Matuszewski_PS}.
We note that not all of the phases must be present. If all three domains are present, we can derive the relation 
between $n_1$ and $n_2$ by calculating the change of the total energy when changing the domain sizes $a$, $b$ and $c$.
Since the magnetization is conserved, we have $\Delta M = b \Delta n_2 - n_2\Delta a = 0$ and consequently
$\Delta n_2 / n_2 = \Delta a / b$. The change of energy is
\begin{align}
\Delta E_{\rm A} &= \left( - \delta E n_1 - \frac{c_2}{2}n_2^2 \right) \Delta a + \frac{c_2}{2} \Delta \!\left(n_2^2\right) = \nonumber\\
&= \left( - \delta E n_1 + \frac{c_2}{2}n_2^2 \right) \Delta a.
\end{align}
This gives the equilibrium condition $\delta E n_1 = c_2 n_2^2/2$, and consequently the necessary (but not sufficient) requirement 
\begin{align}\label{eq_3p}
\delta E < \frac{c_2}{2} n_{\rm max}
\end{align}
for the coexistence of all three phases, where $n_{\rm max}$ is the maximum density in the center of the trap.

\subsection{Ferromagnetic case ($c_2 <0$)}

In the case of a ferromagnetic condensate, the possible ground state phases are $\rho_+$,  $\rho_-$, $\rho_0$, 
and PM~\cite{Matuszewski_PS}. Since it is difficult to propose a simple analytical formula for the energy of the PM phase,
we resort to a qualitative analysis using the condition~(\ref{dem}). 

\begin{figure}
\includegraphics[width=8.5cm]{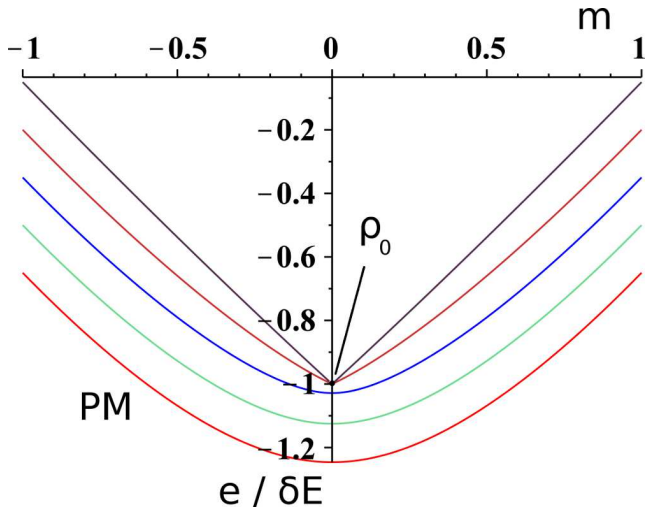}
\caption{Normalized energy per atom versus magnetization for the possible ground states of the ferromagnetic condensate. 
The lines correspond to normalized densities (from top to bottom) $|c_2| n / \delta E = 0.1, 0.4, 0.7, 1.0, 1.3$. 
All the paths fulfilling the condition $d e / d m = \mathrm{const}$,
with the exception of completely magnetized states, arrive at the central
point corresponding to the $\rho_0$ state in the low $n$ limit.
}
\label{e_ferro}
\end{figure}
In Fig.~\ref{e_ferro} we show the normalized energy per atom $e/\delta E$ of these states
in function of the local magnetization $m$ for several values of the normalized density $c_2 n / \delta E$. 
The central point corresponds to the $\rho_0$ phase, while the $\rho_+$ and $\rho_-$ phases correspond to the limits $m=\pm 1$.
The possible Thomas-Fermi profiles correspond to sets of points in the $e,m$ plane that fulfill
the condition $\partial e / \partial m = {\rm const}$. At constant $\delta E$, the magnitude of this derivative generally increases
with decreasing $n$ (when moving from the center of the trap to the less populated regions), 
which can be seen from the gradually ``steeper'' shape of the top curves. To keep
$\partial e / \partial m$ constant, the magnitude of magnetization has to decrease, and eventually we arrive at the 
point $\rho_0$ at a low enough $n$. In this way, we obtain a phase separated state consisting of 
PM and $\rho_0$ domains. The only exceptions from this are the purely $\rho_0$ state for $M=0$ and high enough $\delta E$,
and purely $\rho_+$ and $\rho_-$ states for $M=\pm 1$.

The placement of domains in the PM+$\rho_0$ state can be deduced in a similar way as in the antiferromagnetic case.
The energy per atom in the $\rho_0$ phase is constant, while the energy in the PM phase is a decreasing function of density,
due to the spin-dependent atomic interaction with a negative $c_2$.
Hence the atoms in the PM state will reside in the trap center, where the energy is the highest.

\section{Numerical results} \label{Sec_Numerical}

\subsection{Potential induced spatial separation}

\begin{figure}
\includegraphics[width=8.5cm]{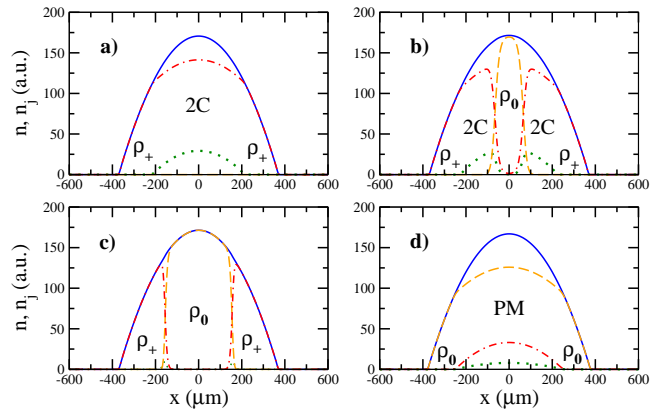}
\caption{Examples of ground states of an antiferromagnetic (a-c) and a ferromagnetic (d) condensate in the quasi-1D case. 
(a) $\mathrm{2C} + \rho_+$ state for $m=0.8$ and $\delta E /c_2 n_{\rm max}= 0.05$,
(b) $\rho_0 + \mathrm{2C} + \rho_+$ state for $m=0.6$ and $\delta E /c_2 n_{\rm max}= 0.17$,
(c) $\rho_0 + \rho_+$ state for $m=0.4$ and $\delta E /c_2 n_{\rm max}= 0.37$, and 
(d) $\mathrm{PM} + \rho_0$ state for $m=0.1$ and $\delta E /c_2 n_{\rm max}= -1.13$. 
The $n_+$,  $n_0$, and  $n_-$ components are depicted
by dash-dotted, dashed, and dotted lines, respectively, and the solid lines show the total density. 
Here $\omega_{\perp}=2\pi\times 10^3\,$Hz,
other parameters are $N=8.4 \times 10^3$ $^{23}$Na atoms, $\omega_{z}=2\pi\times 2.5\,$Hz in (a-c)
and $N=8.4 \times 10^3$ $^{87}$Rb atoms, $\omega_{z}=2\pi\times 1.7\,$Hz in (d).
}
\label{profiles}
\end{figure}

In this section we present examples of numerically determined ground state profiles for both antiferromagnetic and 
ferromagnetic condensates in quasi 1D and quasi 2D setups.
The ground state profiles for a quasi-1D condensate were found numerically 
by solving the 1D version of Eqs.~(\ref{GP}) \cite{Matuszewski_PS}
with rescaled interaction coefficients $c_0^{\rm 1D},c_2^{\rm 1D} = (m \omega_\perp) / (2\pi \hbar) c_0,c_2$, 
where $\omega_\perp$ is the transverse trapping frequency.
The Fermi radius of the transverse trapping potential is
smaller than the spin healing length, 
and the nonlinear energy scale is much smaller than the transverse 
trap energy scale, which allows us to reduce the
problem to one spatial dimension \cite{Beata,NPSE}. 
The solutions were found numerically using the normalized gradient flow method \cite{BaoLim},
which is able to find a state that minimizes the total energy $E$ for given $N$ and $M$, and
fulfills the phase matching condition (\ref{PM}). The conventional imaginary time method
is not suitable for this problem since the normalization of the wavefunction effectively imposes
that the chemical potentials $\mu_j$ are equal, thus failing to take into account some of the ground states.
We note that the direction of the magnetic field with respect to the trap orientation does not influence
our results, since only the contact interactions are taken into account in our model.

\begin{figure*}
\includegraphics[width=16cm]{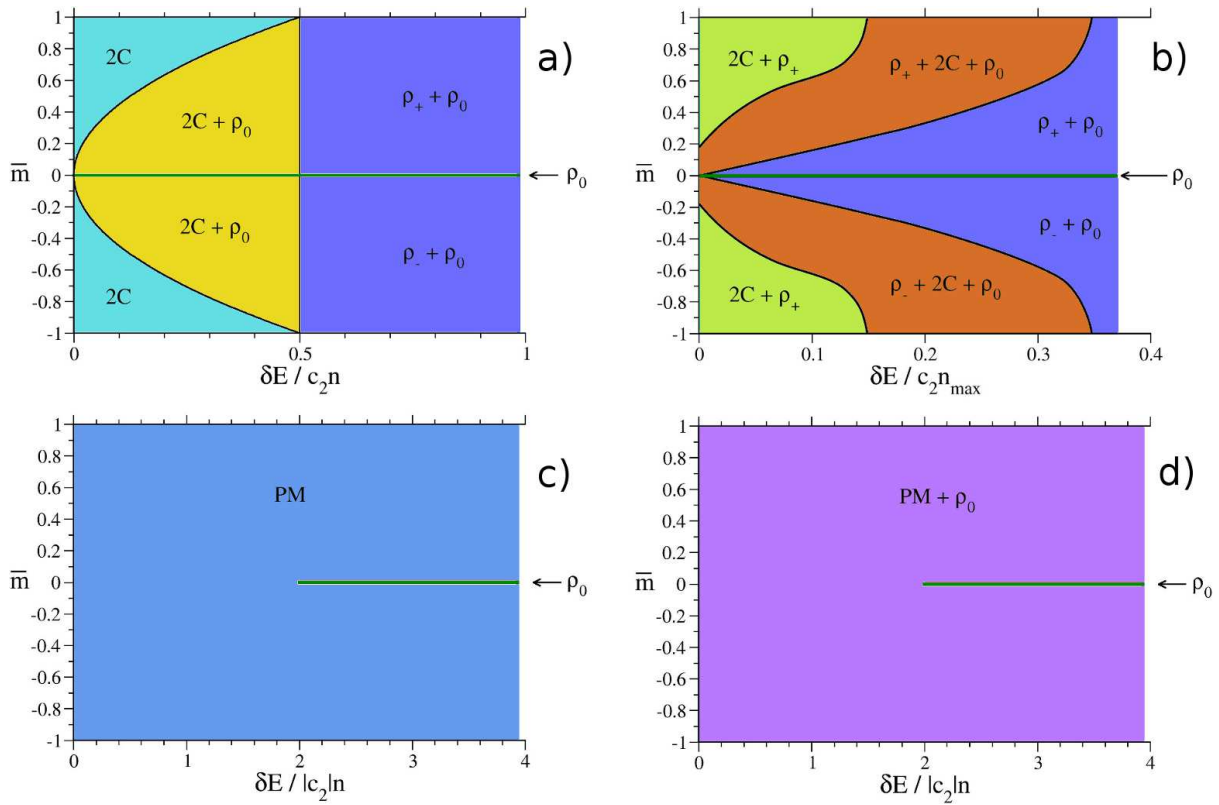}
\caption{Diagrams of phase separation of an antiferromagnetic (a,b) and a ferromagnetic (c,d) condensate
in the quasi-1D case. The left column corresponds to the untrapped case (spontaneous phase separation), 
while the right column corresponds to a condensate in a harmonic potential.
}
\label{phase_diagrams}
\end{figure*}

The setup consists of a highly elongated cigar-shaped harmonic trap, the longitudinal size of the condensate being
much larger than other length scales, i.e.~the spin and magnetic healing lengths $\xi_s$ and $\xi_B$.
This ensures that the contribution of the kinetic energy (e.g.~at the boundaries between domains) is small compared to 
other energy scales in the system, which is the requirement for the application of the Thomas-Fermi approximation.
Figs.~\ref{profiles}(a)-(c) show typical ground state profiles of an antiferromagnetic condensate in various
regimes of parameters. In accordance with the theoretical considerations, the domain structure, in general, consists 
of spatially separated $\rho_0$, 2C, and $\rho_+$ phases, and the arrangement of domains is such that the 
more magnetized the phase, the farther it is from the trap center. In the 2C phase, the difference between the density
of $m=+1$ and $m=-1$ atoms is constant, in agreement with the condition~(\ref{2C_cond}). 

As a rule, the 2C+$\rho_+$  state (a) is the ground state at low magnetic fields, the three-phase state (b) occurs in
the transitory regime, and the $\rho_0$+$\rho_+$ state (c) appears at stronger magnetic fields. This can be understood by
noting that the Zeeman energy of the $m=0$ component is decreased with respect to the $m=\pm 1$ 
components when increasing the magnetic field 
strength. This is the reason for the appearance of the $\rho_0$ domain which eventually replaces the 2C phase completely,
in accordance with the condition~(\ref{eq_3p}). Another possibility is the unmagnetized condensate ($M=0$), 
where the whole condensate consists of a single $\rho_0$ domain (not shown).

In the case of a ferromagnetic condensate, the ground state generally consists of PM and $\rho_0$ domains, as depicted in
Fig.~\ref{profiles}(d). The magnetized PM phase is placed at the center, again in agreement with the results of Sec.~\ref{Sec_TF}.
At low magnetic fields the $\rho_0$ domains can become very small and eventually disappear when their size
becomes smaller than the spin healing length. In this case, the condensate is completely composed of the PM phase.
In a unmagnetized condensate ($M=0$), the phase separated PM+$\rho_0$ state is preferred
at low magnetic fields, $\delta E < 2 |c_2|n$~\cite{Matuszewski_PS}, otherwise a single $\rho_0$ domain is present.

In Fig.~\ref{phase_diagrams}, we summarize the above results in a systematic way and juxtapose them with the ones obtained 
in the case of an untrapped condensate~\cite{Matuszewski_PS}. It is clear that the phase separation
has a qualitatively different character in the presence of the trapping potential. Although there are similarities,
in general the ground states have a different structure. In particular, the antiferromagnetic condensate can separate 
into three distinct phases in the presence of the trapping potential, while the maximum number of phases is two
in the untrapped phase. In the case of a ferromagnetic condensate, the potential induces separation into two phases,
while the untrapped condensate is always homogeneous.

\begin{figure}
\includegraphics[width=8.5cm]{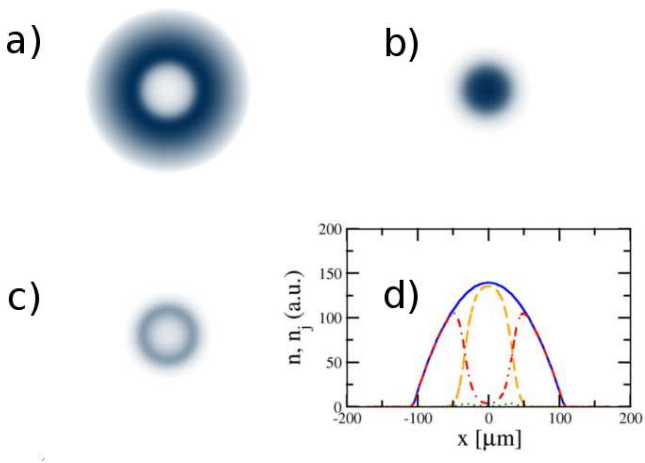}
\caption{The $\rho_0 + \rho_+$ state in a quasi-2D harmonic trap. The densities of the (a) $m=+1$, (b) $m=0$ and (c) $m=-1$ components
are shown, together with the cross section of the densities along the $x$ axis (d). The parameters are $N=2.5\times 10^6$ $^{23}$Na
atoms, $m=0.8$, $B=0.15\,$G, $\delta E /c_2 n_{\rm max}= 0.15$, $\omega_{xy}=2 \pi \times 10\,$Hz,  $\omega_{z}=2 \pi \times 10^3\,$Hz.
}
\label{profiles_2d}
\end{figure}

In Fig.~\ref{profiles_2d} we present an example of a phase separated $\rho_0$+$\rho_+$ ground state of an antiferromagnetic condensate
in the two dimensional geometry. The domains arrange radially, with the unmagnetized phase in the center, 
in agreement with the results of Sec.~\ref{Sec_TF}. Atoms of the $m=-1$ component are present only at the boundary
between the two domains. We note that in the 2D geometry it would be difficult to obtain a clear three phase state, 
due to the limited number of atoms available in experiment.

\subsection{Crossover between the Thomas-Fermi and the single-mode regime}

\begin{figure}
\includegraphics[width=8.5cm]{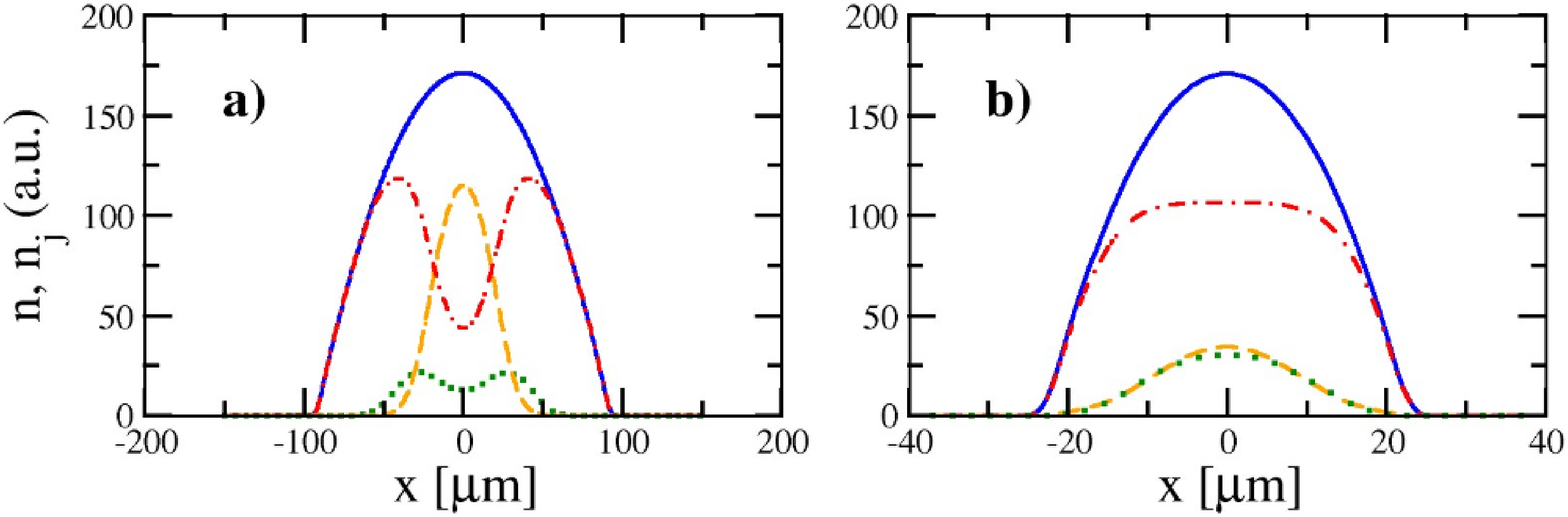}
\caption{Crossover from the Thomas-Fermi regime to the single mode regime and the disappearance of spin domains.
The parameters are the same as in Fig.~\ref{profiles}(b), except that the number of atoms was decreased
(a) $4\times$ and (b) $16\times$, while the trap frequency was increased to keep the density constant.
Here the spin healing length is $\xi_s=17 \mu$m and the magnetic healing length is $\xi_B=42 \mu$m.
}
\label{profiles_tfsma}
\end{figure}

The above results, corresponding to the Thomas-Fermi (TF) regime, are characteristic for the case when the size of the condensate
is much larger than the spin healing length $\xi_s$ and the magnetic healing length $\xi_B$. On the other hand, in the 
opposite limit of a small condensate, the ground state structure is well described by the single mode approximation (SMA)~\cite{SMA},
where all three magnetic components share the same spatial profile. However, typical experimental conditions may correspond
to an intermediate regime, where condensate size is comparable to $\xi_s$ or $\xi_B$. An example of the crossover
between TF and SMA regimes is presented in Fig.~\ref{profiles_tfsma}. The parameters correspond to the three-phase state
from Fig.~\ref{profiles}(b), but the number of atoms was decreased $4\times$ and $16\times$ in (a) and (b), 
while the trap frequency was increased to keep the density constant. The domain structure transforms gradually towards a
single-mode ground state. The thickness of the boundaries between domains is determined by the energy difference between
neighboring phases. In particular, the boundary between 2C and $\rho_+$ domains is of the order of $\xi_s$, while the boundary
between $\rho_0$ and 2C domains is significantly larger, of the order of $\xi_B$. This is clearly visible in 
Fig.~\ref{profiles_tfsma}(b), where the size of the condensate is about two times larger than $\xi_s$ but already smaller than
$\xi_B$. in this case, small $\rho_+$ domains can still be distinguished, but 2C and $\rho_0$ domains merged into a single 
APM phase domain, which is the ground state in the SMA limit~\cite{Matuszewski_PS,Zhang_NJP_2003}.

\section{Generation of spin domains} \label{Sec_Generation}

\begin{figure}
\includegraphics[width=8.5cm]{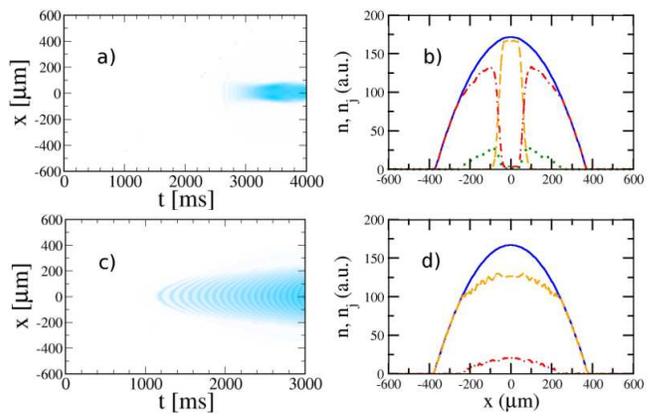}
\caption{
Generation of the ground state profiles by adiabatically changing the magnetic field. The magnetic field is gradually increased from 
zero to the final value $B= 0.15\,$G during $t=4\,$s for sodium condensate (a,b), and 
decreased from $B= 0.5\,$G to $B= 0.35\,$G during $t=3\,$s (c,d) for rubidium. 
The frames (a) and (c) show the time dependence of the atom density in the $m=0$ and $m=+1$ component, respectively, 
and the right column shows the final domain profiles. The parameters are as in Fig.~\ref{profiles}, except
$m=0.6$  in (a,b) and $m=0$ in (c,d).
}
\label{generation}
\end{figure}

From the practical point of view, it is important to propose a reliable method for the generation of the above ground states in experiment.
We show that the method of adiabatic change of the magnetic field, introduced in~\cite{Matuszewski_FNT}, can be used for this purpose.
In the case of an antiferromagnetic condensate, we start from the initial ground state where all the atoms are in the $m=-1$ sublevel.
Next, part of the atoms is transferred to the $m=+1$ sublevel~\cite{Transfer}, and the magnetic field is switched off. In this way,
we obtain a system close to the zero-field ground state with an arbitrary magnetization, determined by the amount of atoms 
transferred. Next, we gradually increase the magnetic field strength in an adiabatic process according to the formula
\begin{equation}
B= \sqrt{\frac{t}{t_{\rm switch}}}\; B_{\rm final} \label{eq_switch}
\end{equation}
where $t=0$ at the beginning of the switching process, $t_{\rm switch}$ is the switching time, 
and $B_{\rm final}$ is the desired final value of the magnetic field.
The form of Eq.~(\ref{eq_switch}) assures that the quadratic Zeeman splitting grows linearly in time. 
In Figs.~\ref{generation}(a,b) we present an example of the generation process simulated by numerical solution of the Gross-Pitaevskii
equations~(\ref{GP}). Fig.~\ref{generation}(a) shows the time dependence of the atom density in the initially 
unoccupied $m=0$ component, and Fig.~\ref{generation}(b) shows 
the final domain profile. This should be compared with the ground state profile in Fig.~\ref{profiles}(b), which corresponds to the
same parameter values.

In the case of a ferromagnetic condensate, the above scenario has to be slightly modified. We begin with strong magnetic field,
in the regime where the quadratic Zeeman energy dominates. The initial state consists of atoms in the $m=0$ sublevel only, which 
is the ground state in these conditions, see Fig.~\ref{phase_diagrams}(d). Here, we are limited to the states with
the total magnetization $M=0$ only. The magnetic field is then decreased, and after crossing the point of phase transition
additional PM domains appear. This scenario and the example of the final state are demonstrated in Figs.~\ref{generation}(c,d).
Since $M=0$, the curves corresponding to densities of $m=-1$ and $m=1$ atoms are overlapping. The PM domain is located in the center of the
trap, where all three components are nonzero.

\section{Conclusions} \label{Sec_conclusions}

We investigated spin-1 Bose-Einstein condensates trapped in a harmonic potentials under the influence of a homogeneous magnetic field.
We demonstrated that the trapping potential has a strong influence on the structure of ground states in these systems,
and can make the antiferromagnetic condensate separate into three, and ferromagnetic condensate into two distinct phases. 
We studied the crossover from the Thomas-Fermi regime to the single mode approximation regime, where the condensate size
becomes smaller than the spin healing length and the spatial structure of ground states disappears.
We suggested an experimental method for creation of these states by adiabatically changing the magnetic field strength.

\acknowledgments

This work was supported by the Foundation for Polish Science through the ``Homing Plus'' programme and by the EU project NAMEQUAM.

\clearpage


\begin{thebibliography}{99}


\bibitem{Ho} T.-L.~Ho, Phys.~Rev.~Lett.~{\bf 81}, 742 (1998); T. Ohmi and K. Machida, J. Phys. Soc. Jpn. {\bf 67}, 1822 (1998).

\bibitem{Entanglement} H. Pu and P. Meystre, Phys.~Rev.~Lett.~{\bf 85}, 3987 (2000);
J. Est\`eve, C. Gross, A. Weller, S. Giovanazzi, and  M. K. Oberthaler, 
Nature {\bf 455}, 1216 (2008).

\bibitem{QS} S. Lloyd, 
Science {\bf 273}, 1073 (1996).

\bibitem{QI} D. DiVincenzo, 
Fortschr. Phys. {\bf 48}, 771 (2000).

\bibitem{Measurement} C. Gross, T. Zibold, E. Nicklas, J. Est\`eve, and  M. K. Oberthaler, 
Nature {\bf 464}, 1165 (2010);
M. F. Riedel, P. B\"ohi, Y. Li, T. W. H\"ansch, A. Sinatra, and  P. Treutlein, 
Nature {\bf 464}, 1170 (2010).

\bibitem{Stenger_Nat_1998} J. Stenger, S. Inouye, D. M. Stamper-Kurn, H.-J. Miesner,
A. P. Chikkatur, and W. Ketterle, Nature (London) {\bf 396}, 345 (1998).

\bibitem{Ketterle_Metastable}
H.-J. Miesner, D. M. Stamper-Kurn, J. Stenger, S. Inouye, A. P. Chikkatur, and W. Ketterle,
Phys. Rev. Lett. {\bf 82}, 2228 (1999).

\bibitem{Sadler_Nat_2006} L. E. Sadler, J. M. Higbie, S. R. Leslie, M. Vengalattore, and D. M. Stamper-Kurn, 
Nature (London) {\bf 443}, 312 (2006).

\bibitem{Mixing} H. Pu, C. K. Law, S. Raghavan, J. H. Eberly, and N. P. Bigelow,
Phys. Rev. A {\bf 60}, 1463 (1999); 
A. T. Black, E. Gomez, L. D. Turner, S. Jung, and P. D. Lett, 
{\it ibid.} {\bf 99}, 070403 (2007).

\bibitem{Chang_PRL_2004} M.-S. Chang, C. D. Hamley, M. D. Barrett, J. A. Sauer,
K. M. Fortier, W. Zhang, L. You, and M. S. Chapman, Phys. Rev. Lett. {\bf 92}, 140403 (2004).

\bibitem{Ketterle_Coreless} A. E. Leanhardt, Y. Shin, D. Kielpinski, D. E. Pritchard, and W. Ketterle, 
Phys.~Rev.~Lett.~{\bf 90}, 140403 (2003).

\bibitem{Zhang_NJP_2003} W.X. Zhang, S. Yi, and L. You, New J. Phys. {\bf 5}, 77 (2003).

\bibitem{Ueda_SBS} K. Murata, H. Saito, and M. Ueda, Phys Rev. A {\bf 75}, 013607 (2007).

\bibitem{SMA} S. Yi, \"O. E. M\"ustecaplioglu, C. P. Sun, and L. You,
Phys. Rev. A {\bf 66}, 011601(R) (2002).

\bibitem{Zhou} F. Zhou, Phys. Rev. Lett. {\bf 87}, 080401 (2001).

\bibitem{Matuszewski_PS} M. Matuszewski, T. J. Alexander, and Y. S. Kivshar,
Phys. Rev. A {\bf 80}, 023602 (2009).

\bibitem{Zhang_PRA_2005} W. Zhang, D. L. Zhou, M. S. Chang, M. S. Chapman, and L. You, Phys. Rev. A {\bf 72}, 013602 (2005).

\bibitem{Timmermans} E. Timmermans, Phys. Rev. Lett. {\bf 81}, 5718 (1998).

\bibitem{Hulet_ImbalancedFermions} G. B. Partridge, W. Li, R. I. Kamar, Y. Liao, and R. G. Hulet
Science {\bf 311}, 503 (2006).

\bibitem{Cornell_BinarySeparation}
D. S. Hall, M. R. Matthews, J. R. Ensher, C. E. Wieman, and E. A. Cornell, 
Phys.~Rev.~Lett.~{\bf 81}, 1539 (1998).

\bibitem{Isoshima_PRA_1999} T.~Isoshima, K.~Machida and T.~Ohmi, Phys.~Rev.~A {\bf 60}, 4857 (1999).

\bibitem{Wuster}  S. W\"uster, T. E. Argue, and C. M. Savage, Phys.~Rev.~A {\bf 72}, 043616 (2005).

\bibitem{Matuszewski_PRA_2008} M. Matuszewski, T. J. Alexander, and Y. S. Kivshar, Phys. Rev. A {\bf 78}, 023632 (2008).

\bibitem{Beata} B.J.~D\c{a}browska-W\"{u}ster, E. A. Ostrovskaya, T. J. Alexander, and Y. S. Kivshar, 
Phys. Rev. A {\bf 75}, 023617 (2007).

\bibitem{Isoshima_PM}
T. Isoshima, K. Machida and T. Ohmi,
J. Phys. Soc. Jpn. {\bf 70}, 1604 (2001); T. Isoshima and K. Machida, Phys. Rev. A {\bf 66}, 023602
(2002).

\bibitem{NPSE} L. Salasnich, A. Parola, and L. Reatto, Phys. Rev. A {\bf 65}, 043614 (2002);
W. Zhang and L. You, Phys. Rev. A {\bf 71}, 025603 (2005).

\bibitem{BaoLim} W. Bao and F. Y. Lim,
SIAM J. Sci. Comput.  {\bf 30}, 1925 (2008); 
F. Y. Lim and W. Bao, Phys. Rev. E {\bf 78}, 066704 (2008).

\bibitem{Matuszewski_FNT} M. Matuszewski, T. J. Alexander, and Y. S. Kivshar,
Fiz. Nizkh. Temp. (in press).

\bibitem{Transfer} J. Kronj\"ager, C. Becker, M. Brinkmann, R. Walser, P. Navez, K. Bongs, and K. Sengstock,
Phys. Rev. A {\bf 72}, 063619 (2005).

\end{thebibliography}
\end{document}